\documentclass{article}

\usepackage{arxiv}

\usepackage[utf8]{inputenc} 
\usepackage[T1]{fontenc}    
\usepackage{hyperref}       
\usepackage{url}            
\usepackage{booktabs}       
\usepackage{amsfonts}       
\usepackage{nicefrac}       
\usepackage{microtype}      
\usepackage{lipsum}
\usepackage{cite}
\usepackage{amsmath,amsfonts,amssymb}
\usepackage{graphicx}
\usepackage{setspace}
\usepackage{tocloft}
\usepackage{hhline}
\usepackage{lineno}
\pdfoutput=1

\usepackage{graphicx}
\graphicspath{{./Figures/}}
\DeclareGraphicsExtensions{.pdf,.jpg,.png}

\usepackage{mathrsfs,amsmath}
\usepackage{mathtools}
\usepackage{algorithm}
\usepackage{algpseudocode}

\usepackage{amsfonts}
\usepackage{tabularx}
\usepackage{multicol}
\usepackage{multirow}
\usepackage{array}
\usepackage{enumitem}
\usepackage{placeins}
\usepackage{hhline}
\usepackage[bottom]{footmisc}
\usepackage{soul,xcolor}
\usepackage{float}
\usepackage{textcomp}

\title{Rapid tissue oxygenation mapping from snapshot structured-light images with adversarial deep learning}

\author{
  Mason T.~Chen \\
    Department of Biomedical Engineering\\
    Johns Hopkins University\\
    3400 N. Charles Street, Baltimore, MD, 21218, USA \\
   \And
  Nicholas J. Durr \\
    Department of Biomedical Engineering\\
    Johns Hopkins University\\
    3400 N. Charles Street, Baltimore, MD, 21218, USA\\
  \texttt{ndurr@jhu.edu}
}

\begin{document}
\maketitle

\begin{abstract}
Spatial frequency domain imaging (SFDI) is a powerful technique for mapping tissue oxygen saturation over a wide field of view. However, current SFDI methods either require a sequence of several images with different illumination patterns or, in the case of single snapshot optical properties (SSOP), introduce artifacts and sacrifice accuracy. To avoid this tradeoff, we introduce OxyGAN: a data-driven, content-aware method to estimate tissue oxygenation directly from single structured light images using end-to-end generative adversarial networks. Conventional SFDI is used to obtain ground truth tissue oxygenation maps for \textit{ex vivo} human esophagi, \textit{in vivo} hands and feet, and an \textit{in vivo} pig colon sample under 659 nm and 851 nm sinusoidal illumination. We benchmark OxyGAN by comparing to SSOP and to a two-step hybrid technique that uses a previously-developed deep learning model to predict optical properties followed by a physical model to calculate tissue oxygenation. When tested on human feet, a cross-validated OxyGAN maps tissue oxygenation with an accuracy of 96.5\%. When applied to sample types not included in the training set, such as human hands and pig colon, OxyGAN achieves a 93.0\% accuracy, demonstrating robustness to various tissue types. On average, OxyGAN outperforms SSOP and a hybrid model in estimating tissue oxygenation by 24.9\% and 24.7\%, respectively. Lastly, we optimize OxyGAN inference so that oxygenation maps are computed ${\sim} 10$ times faster than previous work, enabling video-rate, 25Hz imaging. Due to its rapid acquisition and processing speed, OxyGAN has the potential to enable real-time, high-fidelity tissue oxygenation mapping that may be useful for many clinical applications. 
\end{abstract}


\begin{spacing}{2}   

\section{Introduction}
\label{sect:intro}  
Tissue oxygenation (StO\textsubscript{2}) is a measure of the amount of oxygen in biological tissue, and is often estimated by computing the fraction of oxygenated hemoglobin over total hemoglobin. StO\textsubscript{2} is a useful clinical biomarker for tissue viability, the continuous monitoring of which is valuable for surgical guidance and patient management \cite{edmonds2004cerebral,swartz2003measurements}. Abnormal levels of StO\textsubscript{2} are indicative of many pathological conditions, such as sepsis, diabetes, and chronic obstructive pulmonary disease \cite{sair2001tissue,ditzel1975problem,pitsiou2002tumor}. 

One of the most widely used techniques to measure physiological oxygen levels is pulse oximetry. Despite its ubiquity, robustness, and low cost, pulse oximetry requires a pulsatile arterial signal and only provides a systemic measure of oxygenation \cite{mendelson1992pulse,gioux2011first}. The majority of existing devices for local assessment of StO\textsubscript{2} are based on near-infrared (NIR) spectroscopy. NIR spectroscopy quantifies the concentrations of oxygenated and de-oxygenated hemoglobin by characterizing tissue absorption of light at wavelengths typically between 650 and 1000 nm \cite{siesler2008near}. Similar to pulse oximetry, spectroscopic probes require direct contact with tissue. These measurements can be noisy as they are sensitive to pressure and sample movement \cite{benaron2004continuous,murkin2009near,nguyen2013novel}. Compared to tissue probes, spectroscopic imaging techniques are advantageous as they provide non-contact readings of oxygen saturation at a high spatial resolution and large field of view \cite{zuzak2002visible}. Nevertheless, continuous-wave NIR spectroscopy assumes constant scattering, which could be a source of error as scattering coefficients are often spatially non-uniform. Therefore, to accurately determine oxygen saturation, it is imperative to separate the effect of optical properties, including absorption ($\mu_{a}$) and reduced scattering coefficients ($\mu_{s}'$).

In recent years, spatial frequency domain imaging (SFDI) has emerged as a promising technique for measuring tissue optical properties. SFDI quantifies optical properties by projecting structured light and characterizing the modulation transfer function of tissues in the spatial frequency domain \cite{cuccia2009quantitation}. Oxygenation can subsequently be determined by fitting chromophore concentrations to the measured absorption coefficients using the Beer-Lambert law. In addition to isolating the effect of tissue scattering, SFDI is a wide-field, non-contact technique that can be implemented using a consumer-grade projector and camera. These advantages make SFDI suitable for many clinical applications that necessitate accurate StO\textsubscript{2} measurements, such as burn wound assessment \cite{nguyen2013spatial,burmeister2015utility}, pressure ulcer staging and risk stratification \cite{yafi2017quantitative}, image-guided surgery \cite{gioux2011first, nguyen2013novel}, and cancer therapy evaluation \cite{tabassum2016feasibility}.

\begin{figure}
\centering
\includegraphics[width=\linewidth]{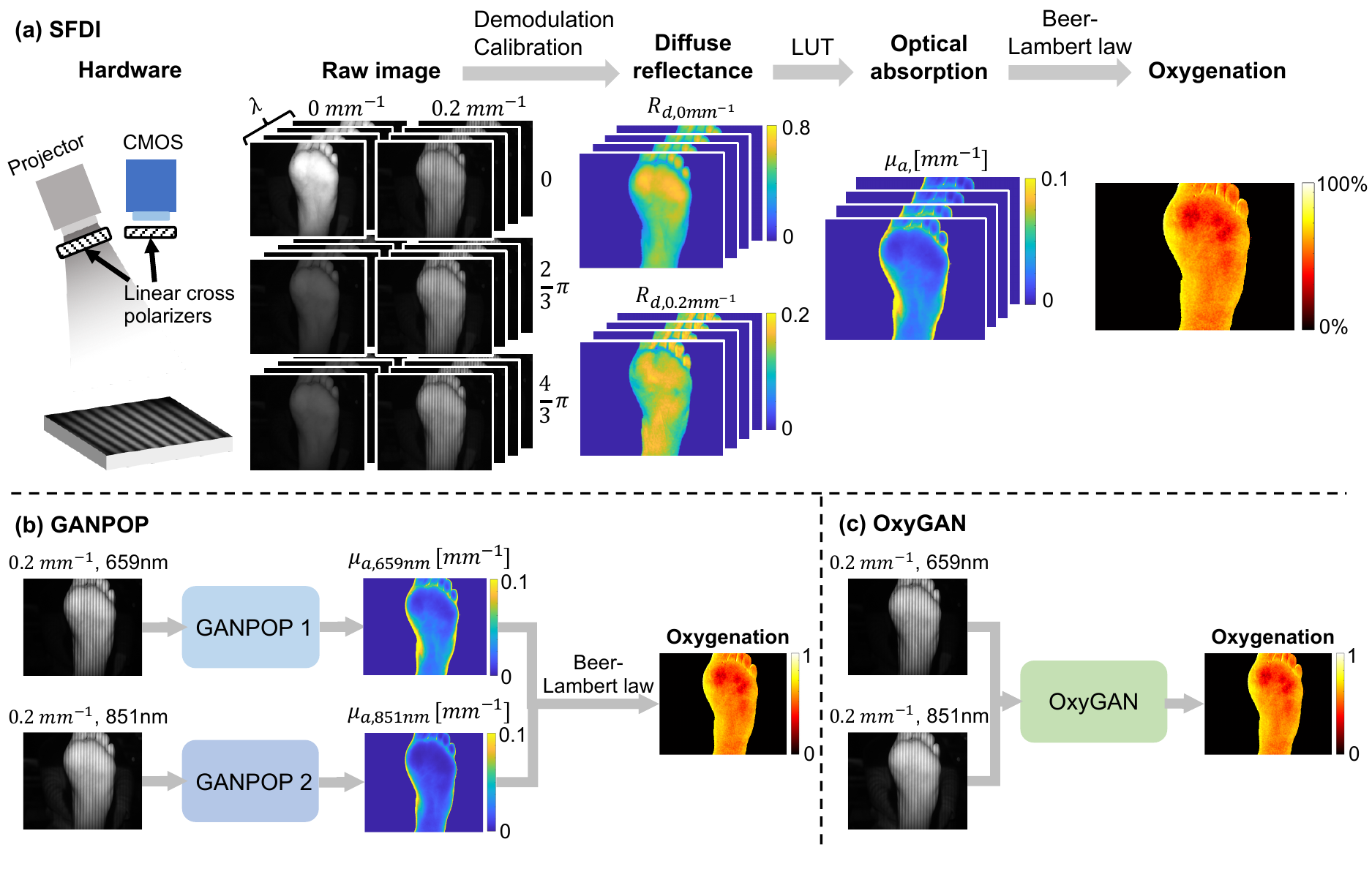}
\caption{Comparison of (a) SFDI, (b) GANPOP, and (c) OxyGAN StO\textsubscript{2} techniques. Ground truth SFDI uses 24 input images (2 spatial frequencies, 3 phases, and 4 wavelengths), while GANPOP and OxyGAN use 2 input images (1 spatial frequency, 1 phase, 2 wavelengths). SFDI and GANPOP absorption maps that are subsequently fit to basis chromophores to estimate StO\textsubscript{2}. OxyGAN directly calculates oxygen saturation with a single network, reducing compounding errors and processing time.}
\label{fig:sfdi}
\end{figure}

Despite its growing use in various applications, there are several factors that limit the clinical translation of SFDI. First, compared to probe-based oximetry, SFDI components are costly. For example, digital micromirror devices or spatial light modulators are often used to produce programmable structured illumination. Second, SFDI requires carefully-controlled imaging geometries, which can be difficult to achieve in clinical settings. Moreover, conventional SFDI requires six images per wavelength and a pixel-wise lookup table (LUT) search to generate a single optical property map. For robust oxygenation estimates, absorption coefficients at a minimum of two wavelengths are needed, and an additional least square fitting step is performed (Fig. \ref{fig:sfdi}(a)) \cite{mazhar2010wavelength}. Previous work has shown that real-time imaging can be achieved with single-snapshot acquisition \cite{schmidt2019real} and either an optimized LUT \cite{angelo2016ultrafast} or a machine learning inversion method \cite{panigrahi2018machine}. However, single image acquisition and frequency filtering often result in image artifacts and high per-pixel error \cite{vervandier2013single}. Therefore, wide-field, rapid, and accurate StO\textsubscript{2} measurement still remains a challenge.

In recent years, convolutional neural networks (CNNs) have emerged as a powerful tool in many medical imaging-related tasks \cite{shin2016deep,shen2017deep}. By employing convolutional filters followed by dimension reduction and rectification, CNNs are capable of extracting high-level features and interpreting spatial structures of input images \cite{ravi2016deep}. For image translation tasks, generative adversarial networks (GANs) improve upon conventional CNNs by utilizing both a generator and a discriminator \cite{goodfellow2014generative} to effectively model a complex loss function. The two components are trained simultaneously, with the generator learning to produce realistic output and the discriminator to classify the generator output as real or fake. Recently, GANs have been employed to predict optical properties from single structured illumination images (GANPOP) \cite{8943974}. As a content-aware network, this technique significantly improves upon the accuracy of model-based single snapshot techniques in estimating optical properties. However, to compute StO\textsubscript{2} with the GANPOP approach, multiple wavelength-specific networks must be run to first estimate absorption coefficients, followed by chromophore fitting, which compounds errors and increases computational demand (Fig. \ref{fig:sfdi}(b)). In this study, we present an end-to-end technique for computing StO\textsubscript{2} directly from structured-illumination images using generative adversarial networks (OxyGAN). OxyGAN maps StO\textsubscript{2} from single snapshot images from 659 and 851 nm sinusoidal illumination. We train generative networks to estimate both uncorrected and profile-corrected StO\textsubscript{2} and compare the performance of the end-to-end architecture versus intermediately calculating optical properties. We accelerate OxyGAN model inference by importing the framework into NVIDIA TensorRT for efficient deployment. Finally, we demonstrate real-time OxyGAN by recording its estimation over the course of a three-minute occlusion experiment.

\section{Methods}

For training and testing of OxyGAN, single structured illumination images were acquired at two different wavelengths (659 and 851 nm) and paired with registered oxygenation maps. Ground truth oxygenation is obtained using the absorption coefficients measured by conventional SFDI at four wavelengths (659, 691, 731, and 851 nm). Experiments were conducted using both profile-corrected and uncorrected ground truth. The training set included human \textit{ex vivo} esophagus samples and \textit{in vivo} feet. OxyGAN was evaluated using unseen tissues of the same type as the training data (human \textit{in vivo} feet) and in different tissue types (human \textit{in vivo} hands and a pig \textit{in vivo} colon). Its performance was additionally compared with single-snapshot optical properties (SSOP) as a model-based benchmark that utilizes a single structured-light image. 

\subsection{Ground Truth Tissue Oxygenation}

In this study, conventional SFDI was used to obtain ground truth StO\textsubscript{2} maps. At each wavelength, structured illumination images were captured using a commercial SFDI system (Reflect RS, Modulim Inc.) at two spatial frequencies (0 and 0.2 mm\textsuperscript{-1}) and three phase offsets ($0$, $\frac{2}{3}\pi$, and $\frac{4}{3}\pi$). The process was implemented for both the sample of interest and a reference phantom. The acquired images were then demodulated and calibrated against the response of the reference at each frequency. The DC (0 mm\textsuperscript{-1}) and AC (0.2 mm\textsuperscript{-1}) diffuse reflectance of the sample were fit to a LUT generated by White Monte Carlo simulations. This pixel-wise LUT search resulted in an optical property map of the sample, which consisted of scattering corrected absorption ($\mu_{a}$) and reduced scattering ($\mu_{s}'$) coefficients. In experiments where profile-corrected ground truth was used, we also implemented height and surface normal angle correction \cite{gioux_three-dimensional_2009,zhao2016angle}. With $\mu_{a}$ measured at four different wavelengths (659, 691, 731, and 851 nm), we subsequently estimated chromophore concentrations using the Beer-Lambert Law:

\begin{equation}
\mu_{a}(\lambda_{i})=\sum_{n=1}^{N}\epsilon_{n}(\lambda_{i})c_{n},
\label{eq:beer}
\end{equation}
where $\epsilon_{n}(\lambda_{i})$ stands for the extinction coefficient of chromophore $n$ at wavelength $\lambda_{i}$, $c_{n}$ is its concentration, and $N$ the total number of chromophores. Oxygen saturation was then calculated as:

\begin{equation}
stO_{2}=\frac{c_{O_{2}Hb}}{c_{O_{2}Hb} + c_{HHb}},
\label{eq:oxygenation}
\end{equation}
where $c_{O_{2}Hb}$ and $c_{HHb}$ represent the concentrations of oxygenated and de-oxygenated hemoglobin, respectively. We estimated ground truth oxygenation maps using absorption coefficients at all the near-infrared wavelengths available in the system (659, 691, 731, and 851 nm). 
    
\subsection{SSOP Benchmark}

In this study, we implemented SSOP as a model-based benchmark. Briefly, this method calculates tissue optical properties from single structured-illumination images by 2-D filtering in the frequency domain. We applied anisotropic low pass filtering using a sine window and high pass filtering using a Blackman window \cite{aguenounon2019single}. The absorption coefficients measured by SSOP at 659nm and 851nm were substituted into Eq. \ref{eq:beer} and \ref{eq:oxygenation} to estimate StO\textsubscript{2}.

\subsection{OxyGAN Framework}

\begin{figure}
\centering
\includegraphics[width=\linewidth]{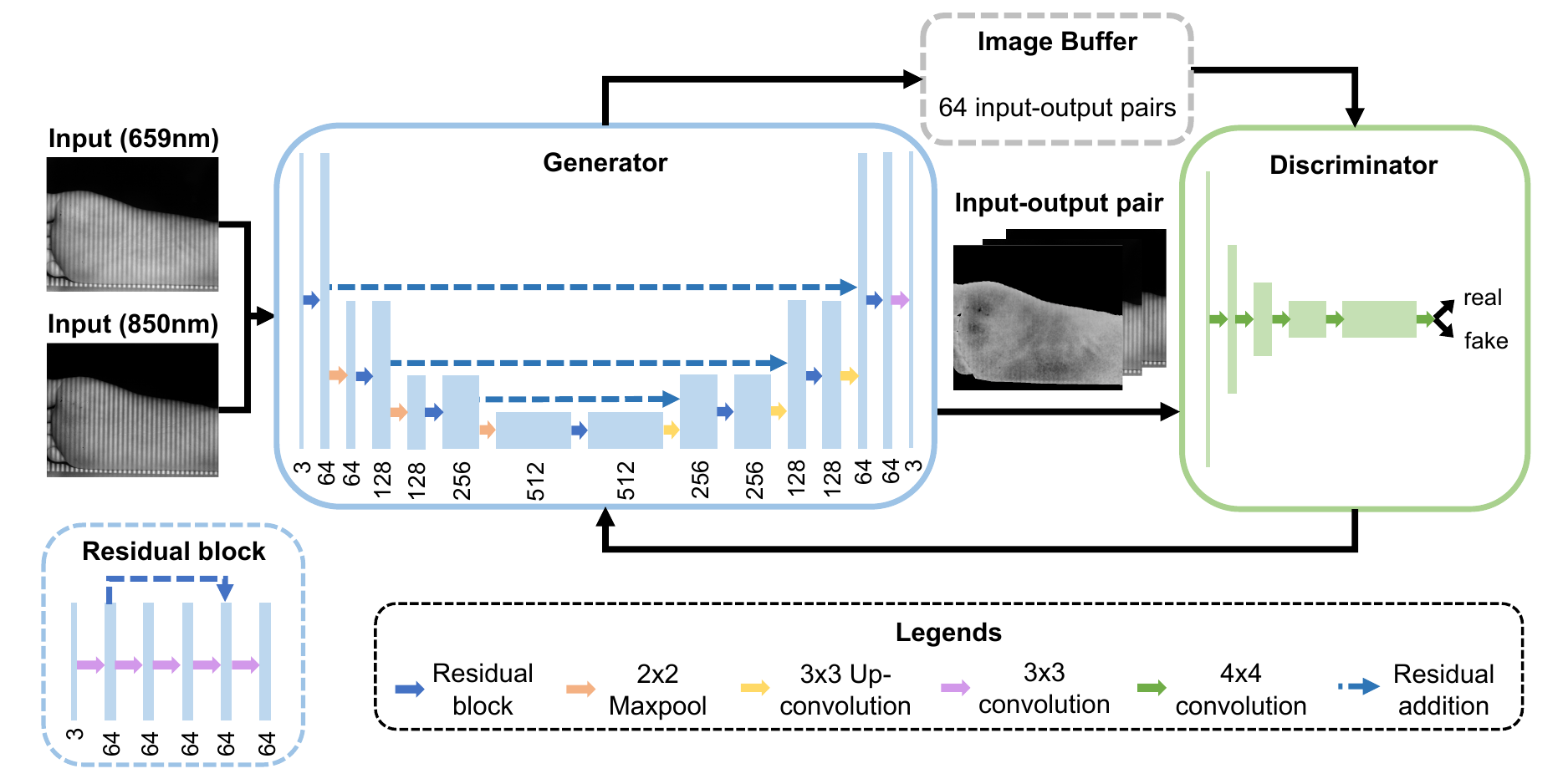}
\caption{The OxyGAN framework. OxyGAN produces StO\textsubscript{2} maps directly from single-phase SFDI images with 659nm and 851nm illumination. The generator is a fusion network that combines the properties of U-Net and ResNet. The number under each block describes the number of channels. The discriminator is a PatchGAN classifier with a receptive field of 70$\times$70 pixels. The discriminator trains to classify the current image pair versus an input-output pair randomly sampled from a pool of 64 previously generated images.}
\label{fig:OxyGAN}
\end{figure}

In this study, we pose StO\textsubscript{2} mapping as an image-to-image translation task. OxyGAN uses an adversarial training framework to accomplish this task (Fig. \ref{fig:OxyGAN}). Specifically, OxyGAN is a conditional generative adversarial network (cGAN) that consists of two convolutional neural networks--a generator and a discriminator. Both networks are conditioned on the same input data, which are single structured-light images in our case. First proposed in \cite{mirza2014conditional}, the cGAN structure has been shown to be an effective solution to a wide range of image-to-image translation problems \cite{isola2017image}. While conventional single-network CNNs require simple, hand-crafted loss functions, cGANs can be more generalizable because the discriminator can effectively learn a complex loss function. 

For the OxyGAN generator, we implement a novel fusion network that combines the properties of U-Net and ResNet (Fig. \ref{fig:OxyGAN}) \cite{Ronneberger_2015,He_2016}. Similar to a U-Net, the OxyGAN generator is an encoder-decoder setup with long skip connections between the two branches on the same level. However, OxyGAN also includes short skip connections within each level and replaces U-Net concatenation with additions, making the network fully residual \cite{quan2016fusionnet,8943974}. The residual blocks on each level consist of five 3$\times$3 convolutional layers, with residual additions between the outputs of the first and the fourth layer. Dimension reduction on the encoder side and expansion on the decoder side is achieved with 2$\times$2 maxpooling and 3$\times$3 up-convolutions, respectively. We use regular ReLUs for the encoder and leaky ReLUs with a slope of 0.2 for the decoder. A final 3$\times$3 convolution followed by a $Tanh$ activation function is applied to generate the output. The discriminator is a three-layer PatchGAN with leaky ReLUs (slope = 0.2) \cite{isola2017image}, which results in a receptive field of 70$\times$70 pixels. To stabilize the training process, we incorporated spectral normalization after each convolution layer in both the generator and the discriminator \cite{miyato2018spectral}. We use use an adversarial loss of:

\begin{equation}
    \mathcal{L}_{\text{cGAN}}(G,D) = \mathbb{E}_{x,y}[\log(D(x,y))] + \mathbb{E}_{x}[\log(1 - D(x,G(x)))],
\label{eq:loss_cgan}
\end{equation}
where $G$ is the generator ($G: X \rightarrow Y$) and $D$ is the discriminator. During training, $G$ tries to minimize this objective while its adversary, $D$, tries to imize it. The discriminator is trained to classify both the current input-ground truth pair and an image pair randomly sampled from a buffer of 64 previously-generated images. Additionally, an $L_1$ loss is included to improve the generator performance and training stability: 

\begin{equation}
    \mathcal{L}_{L1}(G) = \mathbb{E}_{x,y}[||y-G(x)||_1].
\label{eq:loss_l1}
\end{equation}
The full objective function can be expressed as:

\begin{equation}
    \min_{\mathit{G}}\max_{\mathit{D}}\mathcal{L}(G,D) = \mathcal{L}_{\text{cGAN}}(G, D) + \lambda\mathcal{L}_{L1}(G), 
\label{eq:loss_full}  
\end{equation}
where $\lambda$ is a hyperparameter that controls the weight of the $L_1$ loss term and was set to 60. OxyGAN models solved this objective using an \textit{Adam} solver with a batch size of 1 \cite{kingma2014adam}. $G$ and $D$ weights were both initialized from a Gaussian distribution with a mean and standard deviation of 0 and 0.02, respectively. These models were trained for 200 epochs, and a constant learning rate of 0.0002 was used for the first 100 epochs. The learning rate was linearly decreased to 0 for the second half of the training process.  The full algorithm was implemented using Pytorch 1.0 on Ubuntu 16.04 with a single NVIDIA Tesla P100 GPU on Google Cloud \cite{8943974}.

\subsection{Data Split and Augmentation}
\begin{figure}
\centering
\includegraphics[width=\linewidth]{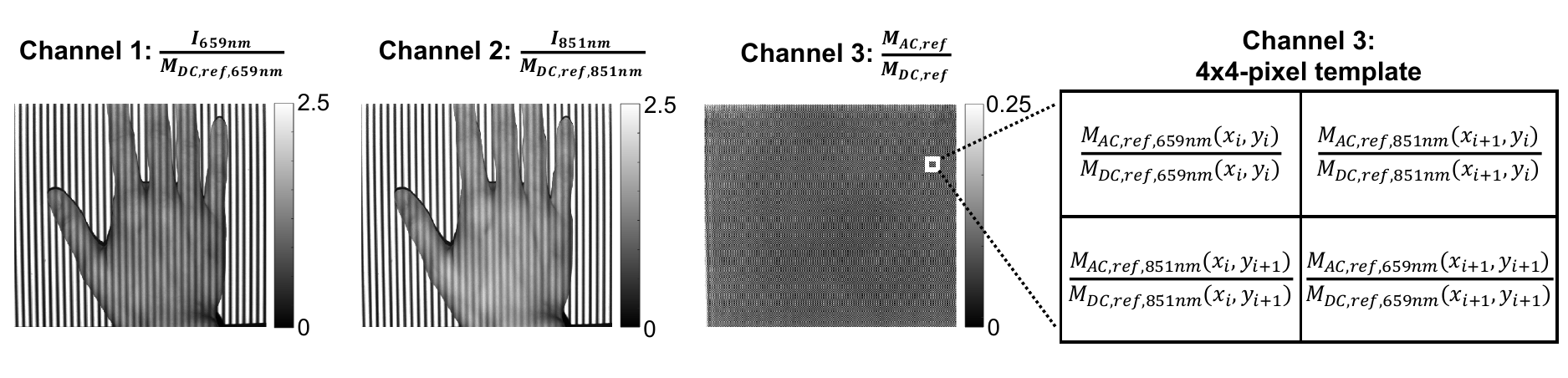}
\caption{Images of the three input channels. Channel 1 and 2 are flat-field corrected single-snapshot SFDI images at 659nm and 851nm. Channel 3 gives ratios between demodulated AC ($M_{AC,ref}$) and DC magnitude ($M_{DC,ref}$) of the reference phantom. The ratios at 659nm and 851nm are alternated to form a checkerboard pattern, as shown in the 4$\times$4-pixel template on the right. }
\label{fig:input}
\end{figure}

In this study, we conducted separate experiments to estimate both uncorrected ($StO_{2}$) and profile-corrected oxygenation ($StO_{2, corr}$) from the same single-snapshot structured light image input. These networks were trained and tested on 256$\times$256-pixel patches paired with registered oxygenation maps. To generate training datasets, each 520$\times$696 image was segmented at a random stride size, which resulted in approximately 30 image pairs per sample. The input data was arranged in a way so that it efficiently utilized the three image channels normally used for color (Fig. \ref{fig:input}). The first and second channel are flat-field corrected, single-phase illumination images at 659 and 851nm, respectively. To account for system drift over time, we included the ratio between demodulated AC ($M_{AC}$) and DC magnitude ($M_{DC}$) of the reference phantom in the third channel. Reference measurements were taken the same day as the tissue measurements, in the same way as conventional SFDI workflows. As shown in Fig. \ref{fig:input}, the ratios at 659 and 851nm alternate in a checkerboard pattern to account for any spatial variations. 

To prevent overfitting of the models, we augmented the training data by flipping the images horizontally or vertically. During each epoch, both flipping operations occurred with a 50\% chance and were independent of each other. Data augmentation was important for this study because of the small size of the training set, and because the testing set includes new object types never seen in training. Additionally, since the classification task of the discriminator was easier than the generator, we applied the one-sided label smoothing technique when training the discriminator. In short, the positive (real) targets with a value of 1 were replaced with a smoothed value (0.9 in our case). This was implemented to prevent the discriminator from becoming overconfident and using only a small set of features when classifying output \cite{salimans2016improved}.

\subsection{Samples}
The training set of OxyGAN models included eight \textit{ex vivo} human esophagectomy samples \cite{sweer2019wide} and four \textit{in vivo} human feet. The testing set consisted of two \textit{in vivo} human hands and feet and an \textit{in vivo} pig colon. All models were cross-validated by training on four of the six feet and testing on the remaining two each time. All summary results reported indicate the average performance of these 3 sets of trained networks. OxyGAN models never see data from hands or \textit{in vivo} pig colon in training.

We additionally recorded a 400 second video of a healthy volunteer's hand during an occlusion study. We first applied a household pressure cuff (Walgreens Manual Inflate Blood Pressure Kit) to the upper arm of the volunteer and imaged the hand at baseline for a minute. Then, the cuff pressure was increased to 200mmHg to occlude the arm for approximately 3 minutes. The pressure was then released and the hand was imaged for another 2.5 minutes. Single-phase sinusoidal illumination was used, which alternated between 659nm and 851nm so that oxygenation could be measured at each time point ($\Delta t = 0.73s$). To obtain ground truth oxygenation, conventional six-image SFDI was implemented every 25 seconds, resulting in 15 measurements in total. 

In this study, the protocols for \textit{in vivo} imaging of human hands and feet (IRB00214149) and \textit{ex vivo} imaging of esophagetomy samples (IRB00144178) were approved by Johns Hopkins Institutional Review Board. The \textit{in vivo} imaging of the pig colon (SW18A164) was approved by Johns Hopkins Animal Care and Use Committee.

\subsection{Performance Evaluation}
In this study, we benchmarked OxyGAN by comparing it to SSOP. We additionally compared OxyGAN to an approach using GANPOP networks to first predict optical properties at 659nm and 851nm and subsequently fitting the concentrations of oxygenated and de-oxygenated hemoglobin using the Beer-Lambert law. These GANPOP networks were trained on the same dataset as OxyGAN with cross validation. The performance of all three methods was evaluated using Normalized Mean Absolute Error (NMAE), which is equivalent to absolute percentage error:

\begin{equation}
NMAE=\frac{\sum_{i=1}^{N}|StO2_{i}-StO2_{i,GT}|}{\sum_{i=1}^{N}StO2_{i,GT}}.
\label{eq:nmae}
\end{equation}
$StO2_{i}$ and $StO2_{i,GT}$ are predicted and SFDI ground-truth oxygen saturation, respectively. $N$ is the total number of pixels. All testing datasets were manually masked to only include pixels that sampled the object.

\section{Results}
The average NMAEs are reported in Table \ref{tab:results} for SSOP, GANPOP, and OxyGAN tested on human feet, hands, and \textit{in vivo} pig colon. The hands and feet are from different healthy volunteers with a wide range of pigmentation levels (Fitzpatrick skin types 1-5). It is worth emphasizing that the \textit{in vivo} hands and pig colon were completely new tissue types that were not represented in the training set. On average, OxyGAN outperforms SSOP and GANPOP in accuracy by 24.04\% and 6.88\%, respectively, compared to uncorrected SFDI ground truth. Compared to profile-corrected ground truth, the improvement of OxyGAN over SSOP and GANPOP becomes 24.89\% and 24.76\%, respectively. 

\begin{table}
\caption{NMAE of StO\textsubscript{2} predicted by SSOP, GANPOP, and OxyGAN on both uncorrected and profile-corrected SFDI ground truth.} 
\label{tab:results}
\begin{center}       
\begin{tabular}{|c||c|c|c|c|c|c|c|c|}
\hline
\rule[-1ex]{0pt}{3.5ex}  & \multicolumn{4}{c|}{\textbf{vs. uncorrected SFDI}} & \multicolumn{4}{c|}{\textbf{vs. profile-corrected SFDI}} \\
\cline{2-9}
\rule[-1ex]{0pt}{3.5ex} & Feet & Hands & Pig & \textit{Overall} &  Feet & Hands & Pig & \textit{Overall} \\
\hline
\rule[-1ex]{0pt}{3.5ex}  SSOP & 0.0396 & 0.0430 & 0.1508 & \textit{0.0778} & 0.0601 & 0.0672 & 0.1404 & \textit{0.0892} \\
\hline
\rule[-1ex]{0pt}{3.5ex}  GANPOP & 0.0438 & 0.0381 & 0.1085 & \textit{0.0635} & 0.0795 & 0.0655 & 0.1220 & \textit{0.0890} \\
\hline
\rule[-1ex]{0pt}{3.5ex}  \textbf{OxyGAN} & \textbf{0.0358} & \textbf{0.0335} & \textbf{0.1080} & \textbf{\textit{0.0591}} & \textbf{0.0536} & \textbf{0.0466} & \textbf{0.1007} & \textbf{\textit{0.0670}} \\
\hline
\end{tabular}
\end{center}
\end{table} 

\begin{figure}
\centering
\includegraphics[width=\linewidth]{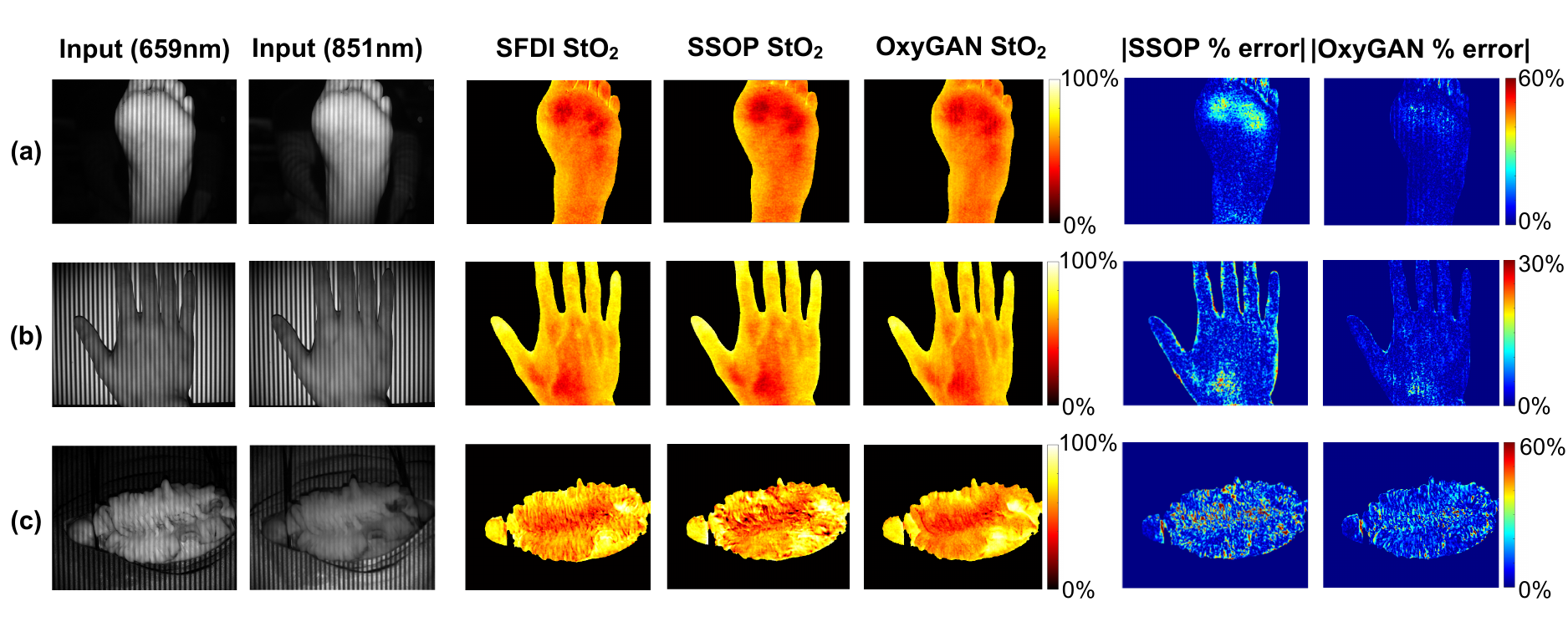}
\caption{Comparison of profile-corrected SFDI, SSOP, and OxyGAN StO\textsubscript{2} results. (a) \textit{in vivo} human foot; (b) \textit{in vivo} human hand; (c) \textit{in vivo} pig colon.}
\label{fig:results}
\end{figure}

Figure \ref{fig:results} compares the results of profile-corrected SFDI, SSOP and OxyGAN applied to a sample of each testing tissue type. Lower errors and fewer image artifacts are observed in OxyGAN results. Error plots highlight the fringe artifacts commonly observed parallel to the illumination patterns in SSOP. As expected, both SSOP and OxyGAN exhibited higher errors in the pig colon, which had more complex surface topography and made single-snapshot predictions more difficult.

\begin{figure}
\centering
\includegraphics[width=\linewidth]{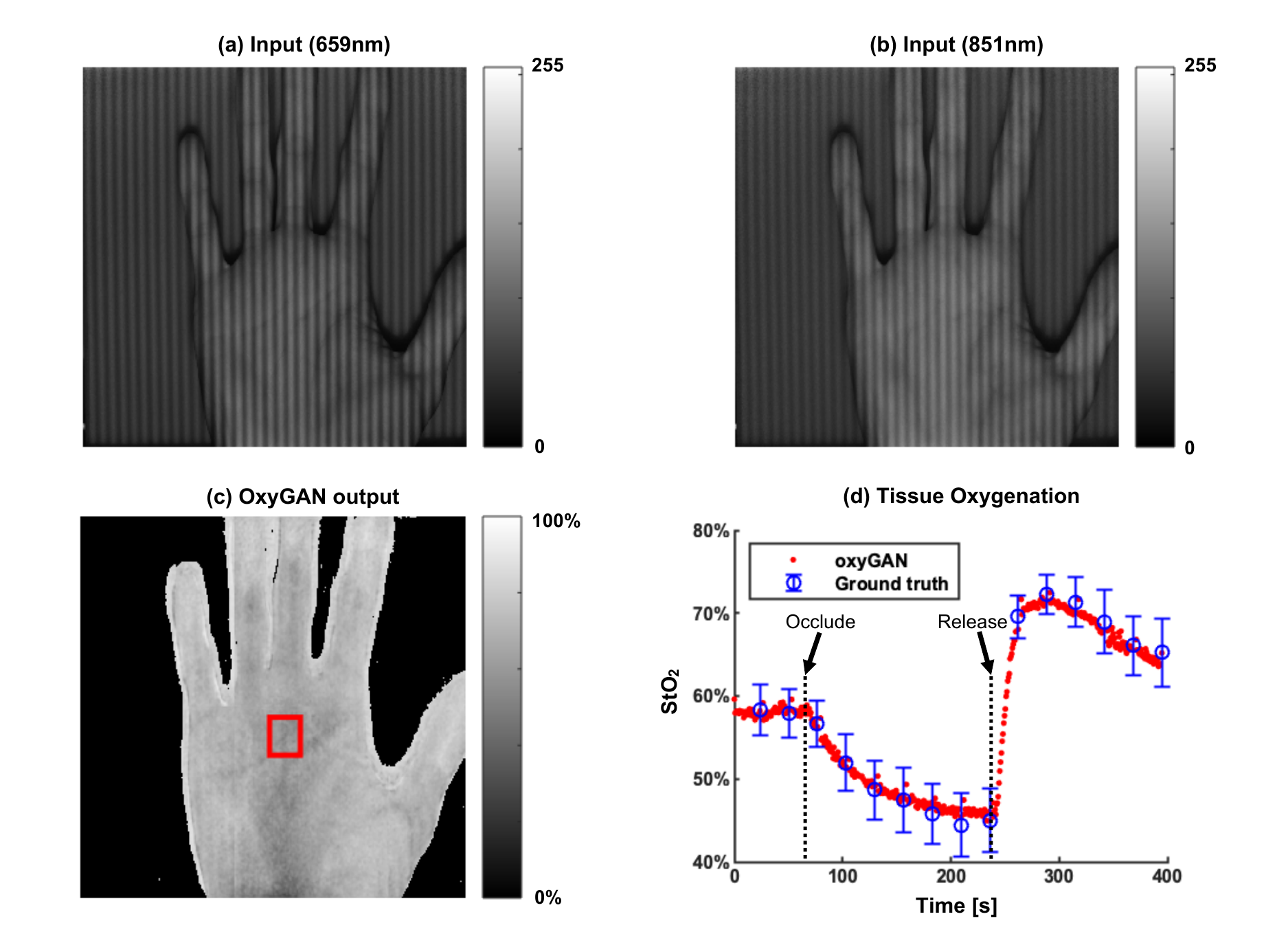}
\caption{Video of the occlusion-release experiment. (a) Input at 659nm; (b) Input at 851nm; (c) StO\textsubscript{2} map predicted by OxyGAN; (d) StO\textsubscript{2} trend measured by OxyGAN and ground truth SFDI over time. (MP4, 3.7MB)}
\label{fig:video}
\end{figure}

We additionally implemented OxyGAN on a video of a volunteer's hand during an occlusion study (Fig. \ref{fig:video}). The average oxygen saturation was calculated for a region of interest highlighted by the red box in Fig. \ref{fig:video}(c) and compared to the SFDI ground truth in Fig. \ref{fig:video}(d). OxyGAN accurately measures a large range of oxygenation values and shows strong and stable agreement with conventional SFDI.

\section{Discussion}
In this study, we have described a fast and accurate technique for estimating wide-field tissue oxygenation from single-snapshot structured illumination images using generative adversarial networks. As shown in Table \ref{tab:results}, OxyGAN accurately measures oxygenation not only for sample types represented in the training set (human feet), but also for unseen sample types (human hands and pig colon). This supports the possibility that OxyGAN can be robust and generalizable. The occlusion video (Fig. \ref{fig:video}) further demonstrates the ability of OxyGAN to accurately measure a wide range of tissue oxygenation levels and detect changes over time. 

Compared to training separate GANPOP networks to first estimate absorption coefficients, OxyGAN produces an average improved accuracy of 15.8\%. Moreover, a greater improvement is observed in profile-corrected experiments. One potential explanation for this is that the errors in absorption coefficients due to uncertainties in profilometry estimation propagate and result in a larger error in oxygenation measurements. Additionally, compared to separate GANPOP models, the end-to-end OxyGAN approach requires only one network and bypasses the Beer-Lambert fitting step, thus greatly reducing the computational cost for training and inference. For example, training a network on 350 patches took approximately 2.2 hours, or 40 seconds per epoch on an NVIDIA Tesla P100 GPU. Training separate GANPOP networks would take double the amount of time and memory. To achieve real-time StO\textsubscript{2} mapping, we first converted the trained model to Open Neural Network Exchange (ONNX) format. We then imported the ONNX model into NVIDIA TensorRT 7 for reduced latency and optimized inference. For testing, OxyGAN inference on a Tesla P100 takes approximately 0.04s to generate a 512$\times$512 oxygenation map. This is 8 times faster than computing optical properties with two GANPOP networks and approximately 10 times faster than two GANPOP inferences followed by a Beer-Lambert fitting step. We achieve We expect OxyGAN to process 1024$\times$1024 images at a similar framerate (25Hz) on a quad-GPU workstation.

To evaluate model performance, we benchmarked OxyGAN by comparing to a single-snapshot technique based on a physical model (SSOP). Table \ref{tab:results} shows that, in estimating both uncorrected and profile-corrected oxygenation, OxyGAN achieves higher accuracy than SSOP in all tissue categories. In addition to improved average accuracy, OxyGAN results also contain fewer subjective image artifacts (Fig. \ref{fig:results}). These benefits are more pronounced for samples with complex surface topography, such as the pig gastrointestinal sample. Unlike SSOP, which relies on Fourier domain filtering, OxyGAN utilizes both local and high-level features. As a content-aware, data-driven approach, OxyGAN has the potential to learn the underlying distribution of the data and accurately infer oxygenation in regions with low signal or non-uniform surface structures.

In Table \ref{tab:results}, we observe that GANPOP achieves similar accuracy to SSOP for profile-corrected ground truth. This is expected for several reasons. First, the training set used in this study is smaller than in the original GANPOP paper \cite{8943974}, excluding \textit{in vivo} hands and tissue-mimicking phantoms. Second, for physical model-based techniques, such as SSOP, the optical property errors due to surface topography variation are correlated across wavelengths, and can later be reduced by chromophore fitting. For instance, for surface normal vectors pointing further away from the detector, the predicted absorption coefficients will be overestimated for both 659nm and 851nm. However, the fitting of hemoglobin concentrations, which relies on the ratios of absorption coefficients, may mask the intermediate optical property errors. Because the GANPOP networks are trained independently for 659nm and 851nm, the loss function does not learn these correlations, resulting in smaller improvements in accuracy over SSOP for StO\textsubscript{2} measurements than for optical property measurements. This observation also provides some intuition for why the OxyGAN network might improve accuracy over GANPOP. Because OxyGAN is trained on multi-wavelength input and the loss function is computed from the StO\textsubscript{2} estimate, it is capable of modeling correlations between absorption at different wavelengths and learning to reduce the effects of varying surface topography.

The architecture of OxyGAN is based on the GANPOP framework \cite{8943974}. The generator combines the features of both the U-Net and the ResNet, in that it incorporates both short and long skip connections and is fully residual. As discussed in Ref.~\citenum{8943974}, this fusion generator has advantages over most other existing architectures because it allows information flow both within and between levels, which is important for the task of optical property prediction. In this study, we empirically trained a model with a standard U-Net generator. The model performed well on sample types included in the training set; however, it collapsed and was unable to produce accurate results when tested on unseen sample types, such as human hands. Compared to GANPOP, OxyGAN employs data augmentation in the form of horizontal and vertical flipping, which is important for preventing overfitting of models trained on small datasets. OxyGAN also utilizes label smoothing in training the discriminator, which further improves model performance and overall training stability. Lastly, we found that adding a channel of checkerboard reference phantom measurements to the 2-wavelength structured light inputs improves accuracy for measurements taken on different days, allowing OxyGAN to take system drift into account similarly to conventional SFDI. 

In the future, more work could be done to optimize the algorithm of OxyGAN to further improve the data processing speed. The model could be trained and tested on larger datasets that span a wider range of tissue types or scenarios that might be encountered clinically. Moreover, oxygenation mapping using SFDI structured illumination is currently limited because it has a shallow depth of field and requires precisely controlled imaging geometry, making its clinical adoption particularly challenging. One alternative is to use random laser speckle patterns as structured illumination, which could be less costly than SFDI projection systems, more easily incorporated into endoscopic applications, and may avoid fringe artifacts due to sinusoidal illumination \cite{chen2020speckle}. Monocular depth estimation could also be incorporated for profile-correction without requiring a projector and profilometry \cite{mahmood2018deep,chen2018rethinking}. Lastly, data-driven methods may be useful for taking higher-order optical property effects into account, such as the scattering phase function.

\section{Conclusion}

In this study, we have presented an end-to-end approach for wide-field tissue oxygenation mapping from single structured illumination images using conditional generative adversarial networks (OxyGAN). Compared to both uncorrected and profile-corrected SFDI ground truth, OxyGAN achieves a higher accuracy than model-based SSOP. It also demonstrates improved accuracy and faster computation than two GANPOP networks that first estimate optical absorption. This technique has the potential to be incorporated into many clinical applications for real-time, accurate tissue oxygenation measurements over a large field of view.

\section{Funding Information}
This work was supported in part by the NIH Trailblazer Award (R21 EB024700).

\bibliography{ref.bib}   
\bibliographystyle{unsrt}

\end{spacing}
\end{document}